\documentclass[a4paper,fleqn,usenatbib]{mnras}

\usepackage[T1]{fontenc}
\usepackage{ae,aecompl}
\usepackage{hyperref}

\usepackage{graphicx}    
\usepackage{amsmath}    
\usepackage{amssymb}    

\usepackage{tikz,xcolor,hyperref}

\definecolor{lime}{HTML}{A6CE39}
\DeclareRobustCommand{\orcidicon}{
	\begin{tikzpicture}
	\draw[lime, fill=lime] (0,0) 
	circle [radius=0.2] 
	node[white] {{\fontfamily{qag}\selectfont \small \hspace{0.05cm}ID}};
	\draw[white, fill=white] (-0.08,0.132) 
	circle [radius=0.0078];
	\end{tikzpicture}
	\hspace{-5mm}
}

\foreach \x in {A, ..., Z}{\expandafter\xdef\csname orcid\x\endcsname{\noexpand\href{https://orcid.org/\csname orcidauthor\x\endcsname}
			{\noexpand\orcidicon}}
}

\title[A source of gamma rays coincident with CTB 80]{A source of gamma rays coincident with the shell of the supernova remnant CTB 80}

\author[Araya \& Herrera]{
  M. Araya\hspace{-3mm}\orcidA{},$^{1}$\thanks{E-mail: miguel.araya@ucr.ac.cr}
  C. Herrera,$^{1}$
  \\
$^{1}$Centro de Investigaciones Espaciales and Escuela de F\'isica, Universidad de Costa Rica}

\date{Accepted ---. Received ---; in original form ---}

\pubyear{2020}

\begin{document}
\label{firstpage}
\pagerange{\pageref{firstpage}--\pageref{lastpage}}
\maketitle

\begin{abstract}
CTB 80 (G69.0+2.7) is a relatively old (50--80 kyr) supernova remnant (SNR) with a complex radio morphology showing three extended radio arms and a radio and X-ray nebula near the location of the pulsar PSR B1951+32. We report on a study of the GeV emission in the region of CTB 80 with \emph{Fermi}-LAT data. An extended source with a size of 1.3$\degr$, matching the size of the infrared shell associated to the SNR, was discovered. The GeV emission, detected up to an energy of $\sim 20$ GeV, is more significant at the location of the northern radio arm where previous observations imply that the SNR shock is interacting with ambient material. Both hadronic and leptonic scenarios can reproduce the multiwavelength data reasonably well. The hadronic cosmic ray energy density required is considerably larger than the local Galactic value and the gamma-ray leptonic emission is mainly due to bremsstrahlung interactions. We conclude that GeV particles are still trapped or accelerated by the SNR producing the observed high-energy emission when interacting with ambient material.
\end{abstract}

\begin{keywords}
gamma rays: ISM -- ISM: individual (CTB 80) -- ISM: supernova remnants
\end{keywords}

\section{Introduction} \label{sec:intro}
CTB 80 was first suggested to be an SNR based on its structure and strong polarization \citep{1974A&A....32..375V}. The radio spectrum of the core was observed to be flat and to steepen away from the core \citep{1981A&A....94..313A}. \cite{1988Natur.331...50K} discovered the 39.5 ms pulsar PSR B1951+32 in the region. This pulsar has a spin-down energy-loss rate $\dot E = 10^{36.6}$ erg s$^{-1}$. \cite{1988Natur.334..229F} discovered a 1$\degr$-diameter infrared shell centered 30$'$ east of the pulsar with an average hydrogen density of 3 cm$^{-3}$. The projected location of this shell suggests that the SNR produced both the shell and the pulsar. Many more observations in radio, the optical and X-rays of the source were carried out following the first studies \citep[e.g.,][]{1988LNP...316..134J,1989MNRAS.237.1109W,1989ApJ...338..171S,1989ApJ...340..362H,2001A&A...371..300M}. \cite{1990ApJ...364..178K} found an H I shell consistent with an SNR shell with a dynamical age of 77 kyr $d_2$, where $d_2$ is the source distance in units of 2 kpc. This shell was found to match the infrared shell. It appears that the pulsar has caught up with the SNR shell, which may have produced the peculiar radio morphology with its interaction with the magnetic field in the shell. The H I shell was later found to contain clumps with core densities of $n_H\sim 100$ cm$^{-3}$ surrounded by a more diffuse envelope (with average density $n_H\sim 1$ cm$^{-3}$), and the interstellar medium (ISM) around CTB 80 was seen to be very inhomogeneous \citep{1993ApJ...417..196K}.  \cite{1995ApJ...439..722S} detected an extended nebula of X-ray emission consistent with synchrotron radiation near PSR B1951+32.

The distance to CTB 80 is still not very clear. For example, based on the dispersion measure of PSR B1951+32 a value of 1.4 kpc has been derived \citep{1988Natur.331...50K}, and more recently, a distance of $4.6\pm0.8$ kpc was derived using red clump stars to measure optical extinction \citep{2018ApJS..238...35S}. The proper motion of the pulsar PSR B1951+32 implies that its age is $64\pm 18$ kyr, assuming a source distance of 2 kpc, according to \cite{2002ApJ...567L.141M}. Later, more precise observations of the pulsar proper motion yielded a kinetic age of 51 kyr \citep[assuming a distance of 2 kpc to the source,][]{2008ApJ...674..271Z}. Based on these studies we adopt a reference distance of 2 kpc for CTB 80 in this work.

High resolution radio observations revealed faint extensions of the arms of CTB 80 and an excellent radio-infrared correspondence along the northern arm \citep{2003AJ....126.2114C}. The ``curling'' observed in the northern arm in radio and infrared may have been caused by the shock interacting with the dense clumps described earlier and this scenario is also consistent with optical observations \citep{2003AJ....126.2114C}. A study of radio spectral index variations along the arms of the SNR revealed a consistent softening of the spectrum away from the pulsar \citep{2005A&A...440..171C}. The spectral steepening was found to be smooth along the eastern arm while the northern and southwestern arms show locally flatter structures, which coincide with optical, radio and infrared enhancements. This feature was interpreted by \cite{2005A&A...440..171C} as a result of the combination of old relativistic electrons injected by PSR B1951+32 and particles accelerated at the sites where the shock of the SNR encounters the inhomogeneities of the ambient medium.

Radio observations by \cite{2012MNRAS.423..718L} revealed an outer slowly-moving H I shell with a radius of 76 arcmin and a velocity of 40 km s$^{-1}$. The shell is consistent with the cool dense shell expected in the ``snowplough'' phase of an SNR. The authors estimate an age for the shell of 60 kyr. They also found extended X-ray emission associated with CTB 80 over a $1^\circ.2$ diameter region. These observations confirmed the large extent of CTB 80 ($\sim 1.2\degr$) implied by observations of \cite{2001A&A...371..300M}.

The average radio spectral index found by \cite{2005A&A...440..171C} is $\alpha=-0.36\pm 0.02$ for the whole SNR ($F_\nu \propto \nu^{\alpha}$), in agreement with the measurements by \cite{1985A&A...145...50M}, while \cite{2006A&A...457.1081K} found $\alpha = -0.45\pm 0.03$. Recent microwave observations by Planck imply the presence of a considerable steepening of the synchrotron spectrum at higher frequencies \citep{2016A&A...586A.134P} which might be caused by electron cooling.

Observations of the CTB 80 region at higher (GeV) energies by the Energetic Gamma Ray Experiment Telescope (EGRET) revealed emission from the pulsar PSR B1951+32 \citep{1995ApJ...447L.109R}, which was later detected by the \emph{Fermi} Large Area Telescope \citep[LAT,][]{2010ApJ...720...26A}. Besides the pulsar, the latest catalog of LAT sources, the Fermi Large Area Telescope Fourth Source Catalog \citep[4FGL,][]{2020ApJS..247...33A}, shows a point source possibly associated to the SNR CTB 80, labeled 4FGL J1955.1+3321. It is found $\sim 0.66^{\circ}$ north east of PSR B1951+32 at the location of the northern radio arm of the SNR shell.

In this work we report the discovery of GeV emission extending across the SNR CTB 80 using data from the \emph{Fermi}-LAT. In section \ref{LAT} we describe the LAT data analysis and in sections \ref{sec3} and \ref{sec4} we present the multi-wavelength data obtained from the literature and the model of the non-thermal emission.

\section{LAT data} \label{LAT}
Data gathered from August 2008 to June 2020 were analyzed with the publicly available software \textit{fermitools} version 1.2.23 and the package \textit{fermipy} version 0.19.0 \citep{2017ICRC...35..824W}. The instrument response functions P8R3\_SOURCE\_V2 were used and standard recommended cuts applied. We selected SOURCE class events in front and back interactions, in the reconstructed energy range 0.3--500 GeV. The maximum zenith angle chosen was 90$^\circ$ to avoid contamination from gamma rays produced in the Earth's limb and time intervals were selected when the data quality was good, filtering events collected while passing the South Atlantic Anomaly and other low-quality events. A spatial binning scale of 0.05$\degr$ per pixel and ten logarithmically spaced bins per decade in energy for exposure calculation were used. Events from a region of interest (ROI) with a radius of 15$\degr$ centred at the coordinates RA=$298.7\degr$, Dec=$33.3\degr$ (J2000) were included in the analysis.

The model of the region included the sources from the 4FGL catalog located within 20$\degr$ of the ROI centre. The source 4FGL J1955.1+3321, possibly associated to CTB 80, was removed from the model to carefully study the emission in the region. The source 4FGL J2005.8+3357, located about $2.4\degr$ from the ROI centre, was also removed and replaced with the extended source associated to 2HWC J2006+341, as described by \cite{Albert_2020}. The Galactic diffuse emission was described by the file {\tt gll\_iem\_v07.fits} and the residual background and extragalactic (isotropic) emission by the file {\tt iso\_P8R3\_SOURCE\_V2\_v1.txt}\footnote{These files are distributed as part of the \textit{fermitools} at https://github.com/fermi-lat/fermitools-data .}. The energy dispersion correction was applied to all components of the model except for the isotropic template. The maximum likelihood technique \citep{1996ApJ...461..396M} was used to fit the free parameters of the model in order to maximize the probability of the model to account for the data. The significance of a new source with one free parameter was estimated with the square root of the test statistic (TS), which is defined as $-2\times$log$(\mathcal{L}_0/\mathcal{L})$, with $\mathcal{L}_0$ and $\mathcal{L}$ the values of the maximum likelihoods for models without the source (the null hypothesis) and with the additional source, respectively. The TS value was also used to choose the best description of a source spectrum from a set of fits with nested functions.

The analysis consisted of a morphological and a spectral characterization of the emission. For the morphological studies, only events with energies above 1 GeV were used in order to take advantage of the improved point spread function of the LAT at higher energies. Only the spectral normalization of the 4FGL sources located within 10$\degr$ of the ROI centre and the normalizations of the diffuse and isotropic components were left free in the fits, while the other spectral parameters were fixed to the values reported in the 4FGL catalog. The null hypothesis was optimized first by searching for new point source candidates in the ROI having a TS$>25$ with the {\tt find\_sources} algorithm of \textit{fermipy}, and using a power-law spectrum. The normalization and spectral index of the new sources were also fitted initially. Different spatial models were used to find the best morphology of the emission in the region of CTB 80, as described below.

Once the morphology of the emission is found the best spectral description of the source was searched using events with energies above 300 MeV. We repeated the search for new point source candidates having a TS$>25$ in this energy range and optimized their normalizations and spectral indices. In the fits, besides the normalizations of the sources located within 10$\degr$ of the ROI centre and those of the diffuse and isotropic components, also the other spectral parameters of the sources located within 2$\degr$ from the centre of the ROI, were left free to vary.

\subsection{Systematic errors}
The effect of the uncertainty in the model of the Galactic diffuse emission on the source parameters was estimated by repeating the fits using the eight alternative models developed by the LAT Collaboration in their search for high-energy emission from supernova remnants \citep{2016ApJS..224....8A}. The files released by the LAT Collaboration were scaled to account for differences in energy dispersion between Pass 7 reprocessed data and Pass 8 data\footnote{See https://fermi.gsfc.nasa.gov/ssc/data/access/lat/Model\_details/\\Pass8\_rescaled\_model.html.}. The uncertainties were estimated as in \cite{2016ApJS..224....8A} for both the spectral parameters of the global fit to CTB 80 as well as for the individual spectral energy distribution (SED) flux points obtained. For the SED points these systematic errors were comparable to the statistical errors below an energy of 4 GeV, while the statistical uncertainties dominated above this energy. The uncertainties in the LAT effective area were propagated onto the global spectral parameters and SED flux normalizations using a set of bracketing response functions as recommended by \cite{2012ApJS..203....4A}. In the fits with the bracketing response functions the pivot energy value, estimated with the covariance error matrix of the global fit, was used.

\subsection{Results}
\subsubsection{Source extension}\label{morphology}
\begin{figure}
 \includegraphics[width=\linewidth]{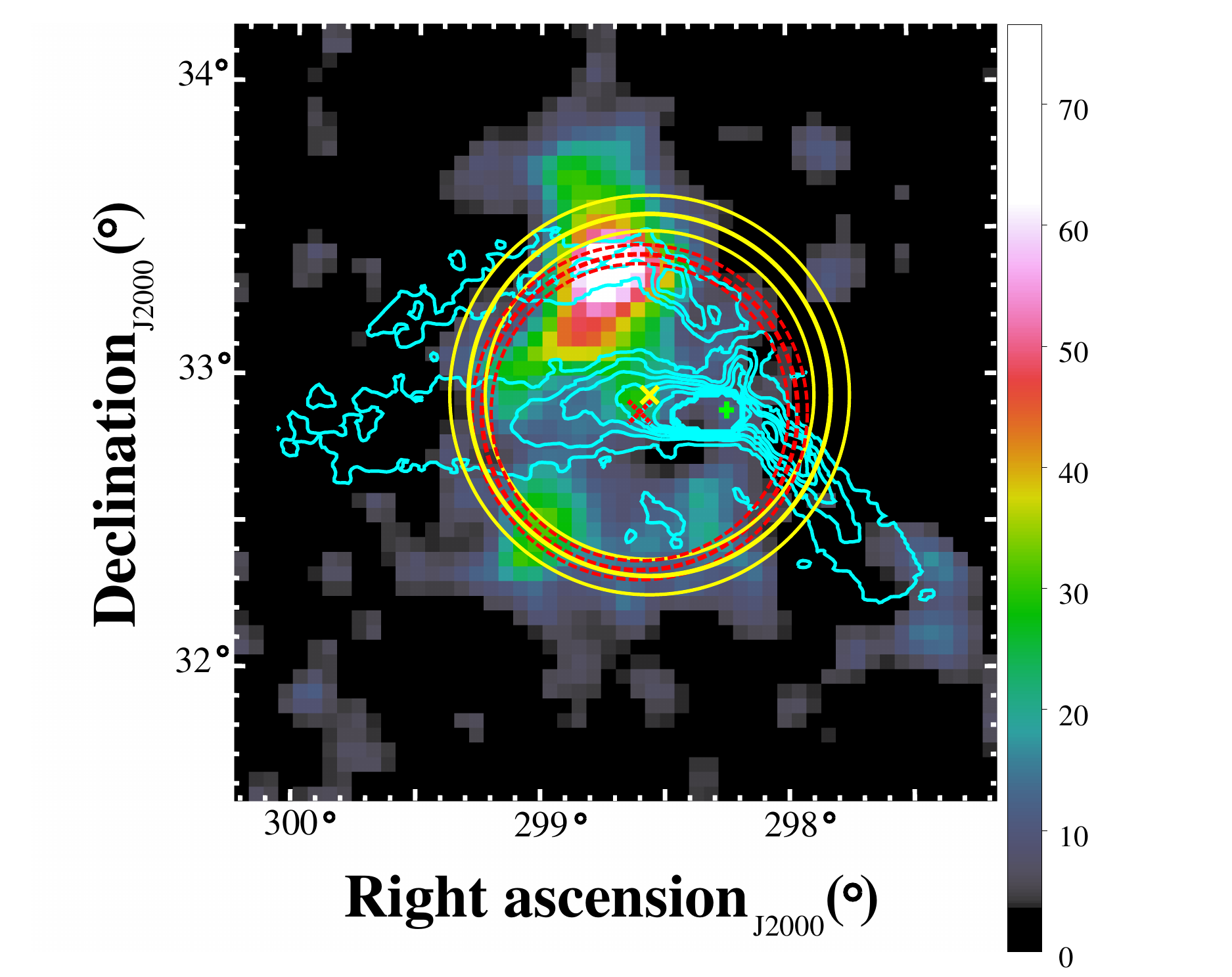}
\caption{TS map of the gamma-ray emission above 1 GeV once the background sources and diffuse emission are subtracted. The color scale is in units of TS. The cross indicates the location of the source 4FGL J1952.9+3252, associated to the pulsar PSR B1951+32. The contours represent the radio emission of CTB 80 in six equally-spaced intervals from 0.002 to 0.202 Jy/beam \citep[taken from a GB6 survey,][]{1994AJ....107.1829C}. The dashed (solid) circle represents the 68\% containment radius and its statistical uncertainty of the best-fit disk (gaussian) template of the GeV source while the X marks the corresponding centroid.}\label{fig1}
\end{figure}

The optimized model for the null hypothesis using events with energy above 1 GeV was used to create a significance map of the ROI which shows the residual emission. A close-up image of the map in the CTB 80 region is shown in Fig. \ref{fig1}. To make the image the TS value of a test point source with a spectral index of 2.0 was evaluated in each spatial bin. The position of 4FGL J1952.9+3252 (included in the model) is marked in the map and corresponds to the pulsar PSR B1951+32, associated to CTB 80. The radio contours plotted in the figure were taken from the 4850 MHz GB6/PMN survey \citep{1994AJ....107.1829C} and show the three characteristic arms of the SNR. The map clearly shows the presence of significant emission at GeV energies, mainly at the location of the northern radio arm but also extending towards the south of CTB 80.

In order to characterize the morphology of the emission several hypotheses were tested in the fits: a point source, a 2D Gaussian template, a uniform disk template and a map of the radio emission. For the radio map, data from the GB6 survey at 4850 MHz was used \citep{1994AJ....107.1829C}. The radio emission from the pulsar wind nebula (PWN) associated to PSR B1951+32 was removed with a $8'\times 4'$ mask, a size given by \cite{2003AJ....126.2114C}, as well as the emission from a point source that is clearly visible at the coordinates RA = $299\degr.05$, Dec = $33\degr.31$ using a mask with a radius that is somewhat larger than the point-source response indicated in the survey ($\sim 1.8'$). Changing the sizes of the masks has no effects on the results shown here. The spectral model used in the fits was a power law whose normalization and spectral index were free to vary. The choice of this spectral shape is justified below. The spatial and spectral parameters of the different models were fitted and Table \ref{table} shows the results. The Akaike Information Criterion \citep[AIC,][]{1974ITAC...19..716A}, defined as AIC = $2k - 2\ln(\mathcal{L})$ where $k$ is the number of parameters and $\mathcal{L}$ the maximum likelihood, was calculated to identify the best model. In the table the value of $\Delta\mbox{AIC} \equiv \mbox{AIC}_k - \mbox{AIC}_{min}$ is given, which is the difference of the AIC of each model $k$ and the one that minimizes the AIC ($\Delta$AIC = 0 for the best available model).

\begin{table}
\caption{Results of the morphological analysis of \emph{Fermi}-LAT data above 1 GeV.}
\label{table}
\begin{center}
\begin{tabular*}{\textwidth}{ccccc}
  \hline
\hline
\textbf{Morphology} & \textbf{Size$^{a}$ ($\degr$)} & \textbf{RA ($\degr$)} & \textbf{Dec ($\degr$)} & \textbf{$\Delta$AIC}\\
\hline

Point source & - & $298.72 \pm 0.03$ & $ 33.36 \pm 0.03$ & 36.9\\

Disk & $0.65 \pm 0.04$ & $298.60\pm 0.04$ & $32.87 \pm  0.04$ & 0\\

Gaussian & $0.41 \pm 0.04$ & $298.56 \pm 0.05 $ & $32.93 \pm 0.05$ & 0.4\\

Radio template & - & - & - & 36.4\\

\hline
\end{tabular*}\\
\textsuperscript{$a$}\footnotesize{Radius for the disk and $\sigma$ for the gaussian.}\\
\end{center}
\end{table}

The Gaussian and disk morphologies provide similarly good representations of the data. The 68\%-containment sizes for both the best-fit disk and Gaussian are shown in Fig. \ref{fig1}, as well as their corresponding 1$\sigma$-statistical uncertainties. Even though these spatial shapes provide the best description of the GeV emission among the models compared, the actual photon distribution is clearly more complex, as can be seen in Fig. \ref{fig1}. Although the residual significance maps obtained after adding the source, for both the disk and Gaussian morphologies, were the most satisfactory among the tested models, a $3.3\sigma$ excess was still seen at the location of the peak of the emission in Fig. \ref{fig1} in both cases. A more detailed study of the morphology should be carried out in the future. The likelihood ratio between the best-fit point source (ps) and the best-fit disk (ext) hypotheses was TS$_{\mbox{ext}} \equiv 2\times$log$(L_{\mbox{ext}}/L_{\mbox{ps}}) =88$. The threshold to prefer the extended source hypothesis over a single point source has been set as TS$_{\mbox{ext}}>16$ \citep{2017ApJ...843..139A}, based on simulations with LAT data showing that the cumulative density of TS$_{\mbox{ext}}$ follows a $\chi^2$ distribution with one degree of freedom \citep{2012ApJ...756....5L}. The emission found is statistically significantly extended. The radius of the best-fit disk found is $0\degr.65 \pm 0\degr.04$. We note that the emission centroids of the best-fit disk and, particularly, the Gaussian, are not located at the position of the PWN (the statistical uncertainties in the position are smaller than the marker sizes in Fig. \ref{fig1}), and there are no other known pulsars in the region. We also note that the disk size and location match well those corresponding to the SNR shell seen in the infrared (see also Fig. \ref{fig3} below). We note, however, that a detailed model of the pulsar emission during its lifetime should be explored and could perhaps explain the GeV morphology as originating from leptons from the pulsar which are now located in the shell of the SNR. In what follows we use the disk template to represent the GeV emission from CTB 80.

\subsubsection{Spectrum}
Using the best-fit disk found in section \ref{morphology}, fits with events in the entire energy range considered (0.3--500 GeV) were done with different spectral shape models. Two models were compared, a power law spectrum, $\frac{dN}{dE}=N_0 (\frac{E}{E_0})^{-\Gamma}$, and a log-parabola, given by $\frac{dN}{dE}=N_0 (\frac{E}{E_b})^{-(\alpha + \beta \log(E/E_b))}$, where $E_0$ and $E_b$ are fixed scales. The difference in log$\mathcal{L}$ between the two models is $\sim 1.6$, and therefore the power law is chosen as the final spectral shape for the GeV emission from CTB 80, since the log-parabola does not improve significantly the fit with respect to the power law. This justifies the use of this spectral shape in the search for the best-fit morphological template above. The overall TS value of the best-fit disk is 349.7, corresponding to a detection significance of $18\sigma$ above 300 MeV.

Setting $E_0 = 10^4$ MeV, the values of the best-fit spectral index and normalization are $\Gamma = 2.34\pm 0.09_{stat}\,\pm 0.08_{sys}$, and $N_0=(1.78 \pm 0.31_{stat}\pm 0.17_{sys})\times 10^{-14}$ MeV$^{-1}$ cm$^{-2}$ s$^{-1}$.

\begin{figure}
 \includegraphics[width=\linewidth]{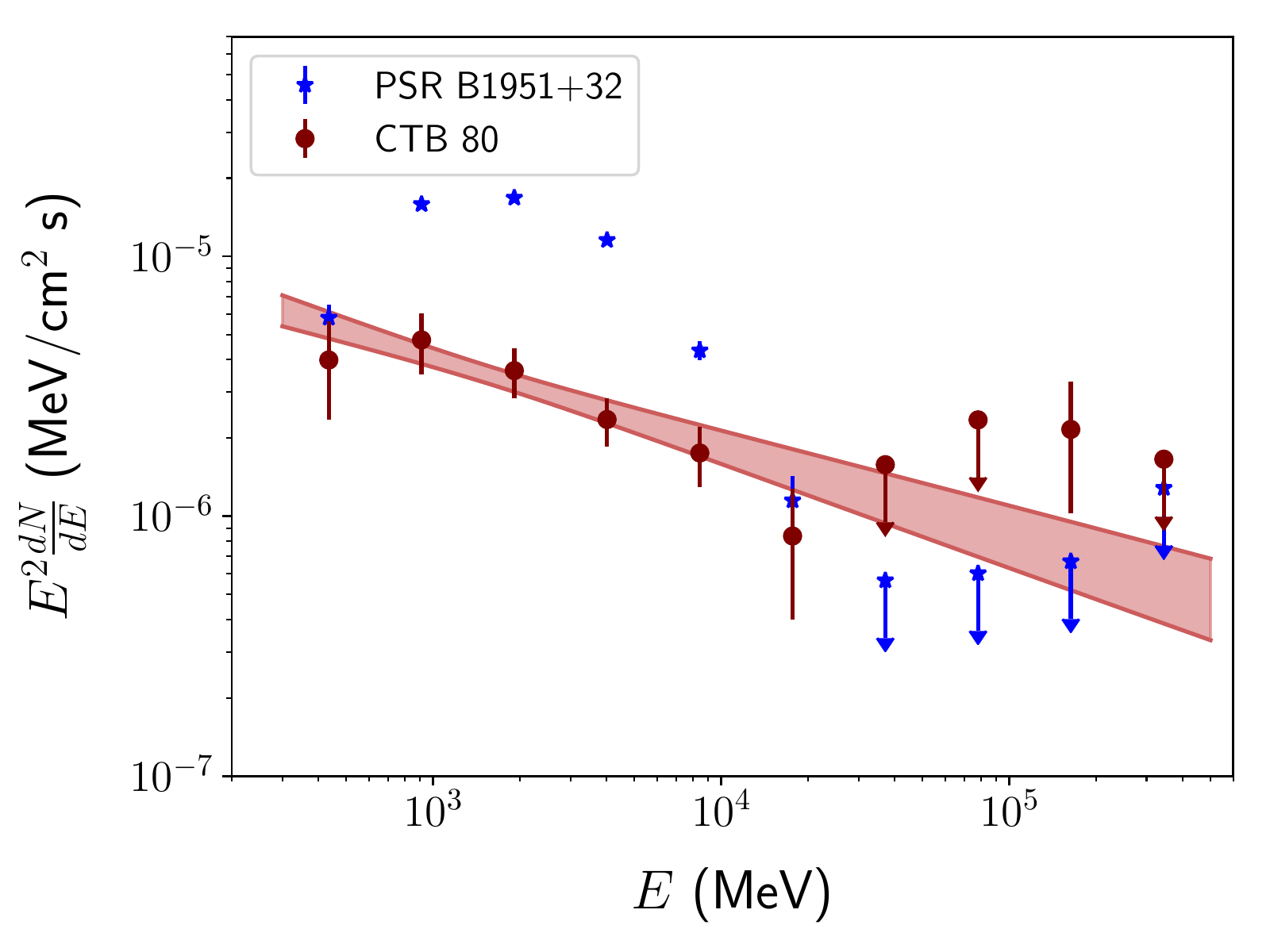}
\caption{Gamma-ray SEDs of the SNR CTB 80 and the pulsar PSR B1951+32 obtained in this work. The circles represent the fluxes from CTB 80 while the stars those from the pulsar. The shaded region represents the propagated $1\sigma$-statistical error band of the global fit for CTB 80. The flux points show statistical and systematic errors added in quadrature.\label{fig2}}
\end{figure}

The SED of the new extended GeV source in the CTB 80 region was obtained by dividing the energy range in 10 logarithmically spaced bins and fitting the normalization of the emission in each bin using the best-fit morphological model obtained. The spectral index value of the disk representing CTB 80 was fixed to 2 in each energy bin and the normalization of the sources located within $2\degr$ of the centre of the ROI (as well as the normalization of the disk and the diffuse and isotropic components) were free to vary. In a similar fashion, the SED points of the pulsar PSR B1951+32 were also obtained for comparison. The resulting SEDs are seen in Fig. \ref{fig2}. The source associated to CTB 80 is not significantly detected above $\sim20$ GeV and 95\%-confidence level upper limits on the fluxes were calculated in the bins where a TS was lower than 4. A TS value of 6.6 was obtained for CTB 80 in the energy bin 113-237 GeV but it is not clear if this is a statistical fluctuation or indication of spectral hardening, and a more detailed study on this is left for the future.

The SED flux points from CTB 80 were calculated in an identical manner replacing the standard diffuse emission model with the eight alternative emission models described earlier and the systematic error on each point was calculated. These errors were added in quadrature to the error resulting from propagation of the effective area uncertainties in the normalization, estimated to be $\sim5$\%, and the resultant errors were also added in quadrature to the statistical uncertainties for each SED point.

\section{Multi-wavelength observations}\label{sec3}
\begin{figure}
 \includegraphics[width=\linewidth]{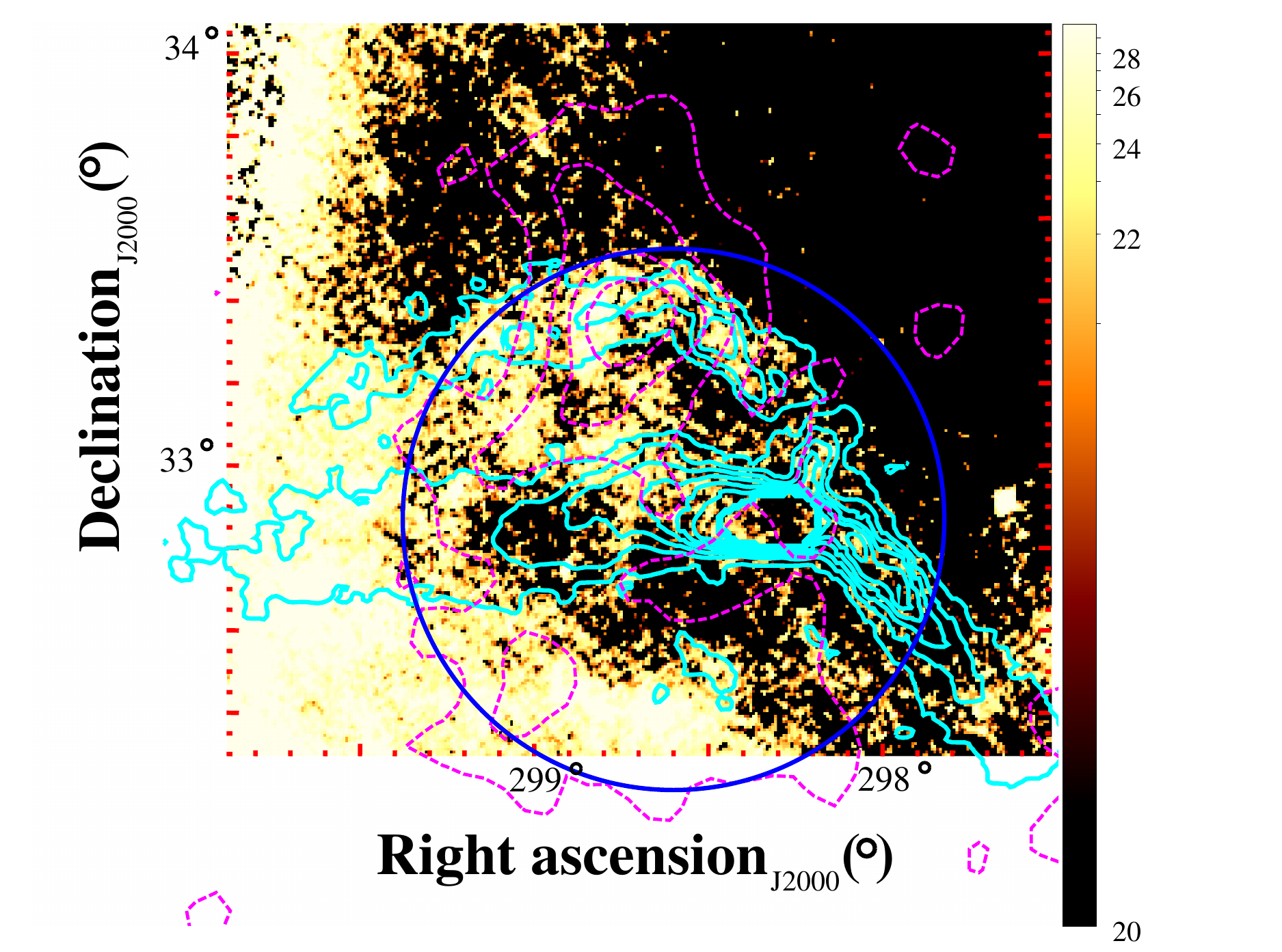}
 \caption{Infrared image of CTB 80 at 65 $\mu$m from the AKARI far-infrared survey. The color scale is in MJy/sr. The radio contours shown in Fig. \ref{fig1} are also shown here (solid, cyan), the (dashed) magenta contours represent the TS values in Fig. \ref{fig1} at the levels of 8, 25, 42, 59 and 76, while the solid (blue) circle represents the disk found at GeV energies. The peak of the GeV emission is located in the northern arm of the SNR where an enhancement of the IR emission is also seen. The circular infrared shell associated to the SNR can be partly seen in this image at the location of its northern radio arm.\label{fig3}}
\end{figure}

Fig. \ref{fig3} shows an infrared image of the CTB 80 region obtained with data from the AKARI far-infrared all-sky survey \citep{2015PASJ...67...50D,2015PASJ...67...51T}. The image partly reveals the infrared shell associated to the SNR \citep{1988Natur.334..229F}. The gamma-ray (TS) contours obtained in this work are plotted as well as the radio contours shown previously. The peak of the GeV emission is seen in the northern arm of the SNR at the location where there is enhancement of the infrared emission (near the coordinates RA =$298.8\degr$ , Dec=$33.3\degr$). This infrared enhancement has been noticed to perfectly match the radio emission and is consistent with shock-heated dust with a temperature $\sim 26$ K \citep{2003AJ....126.2114C}. The interaction of the shock with dense gas could also be responsible for the curving of the northern radio arm. It is also worth noting that the infrared shell partly seen in the image is consistent with the location and size of the gamma-ray disk found here (the circle in Fig. \ref{fig3}).

We used the radio fluxes from CTB 80 from the literature as listed by \cite{2003AJ....126.2114C} and added the more recent Planck measurements in the microwave \citep{2016A&A...586A.134P} which are useful to constrain the highest energy in the electrons. We excluded the radio fluxes below a frequency of 200 MHz from the model, as they could be affected by free-free thermal absorption \citep{2005A&A...440..171C}.

\section{SED modeling and discussion}\label{sec4}
The radio to GeV non-thermal fluxes were modeled using one-zone leptonic or hadronic scenarios with the {\tt naima} package \citep{naima}. For the hadrons we used a particle distribution which is a power law, which is enough to explain the data, while the lepton distribution is described by a power law with an exponential cutoff, where the cutoff defines the maximum attainable energy of the leptons. The leptonic gamma-ray emission is produced by the inverse Compton scattering (IC) and bremsstrahlung processes. For the IC calculation, we used three photon background fields, the cosmic microwave background (CMB) a far-infrared (FIR) photon field and stellar optical and near-infrared (NIR) photons. From the bolometric infrared luminosity of the SNR, $2.2\times10^{37}$ erg s$^{-1}$ \citep{1988Natur.334..229F}, an estimation of the energy density in the FIR photon field at the surface of the shell yields $\sim0.01$ eV cm$^{-3}$. This value is negligible compared to the Galactic FIR energy density at a Galactocentric distance of 8 kpc, $\sim 0.35$ eV cm$^{-3}$ \citep{2011ApJ...727...38S}. We adopt an FIR component produced by a modified black body with a temperature of 26 K and an energy density of $0.35$ eV cm$^{-3}$. The NIR photon field is similarly described with an energy density of 0.7 eV cm$^{-3}$ and a temperature of 2000 K, comparable to the local (Solar System) values \citep{2011ApJ...727...38S}. As will be seen, the results do not depend strongly on these parameters. To calculate the bremsstrahlung and pion decay fluxes, we fixed the ambient target density to 3 cm$^{-3}$, consistent with the average hydrogen density measured in the infrared shell \citep{1988Natur.334..229F}, part of which is seen to coincide with the location of the peak of the gamma rays in Fig. \ref{fig3}.

\subsection{Leptonic model}
\begin{figure}
 \includegraphics[width=\linewidth]{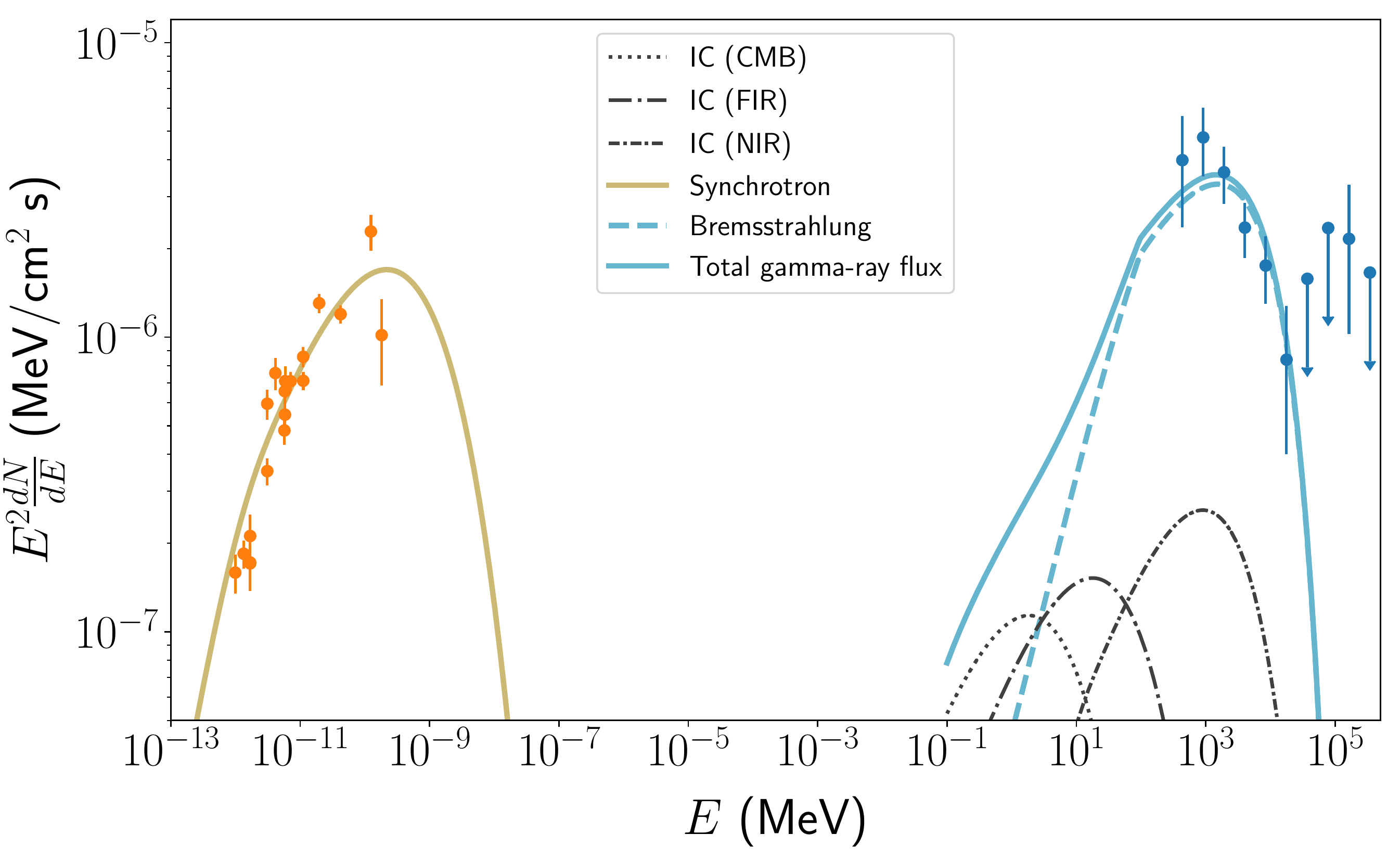}
\caption{Leptonic model for the non-thermal emission from the SNR CTB 80. The gamma-ray data include both statistical and systematic errors. \label{fig4}}
\end{figure}

We fixed the particle spectral index to 1.72 as expected from radio observations \citep{2005A&A...440..171C} and adjusted the other parameters. The resulting model and the data are shown in Fig. \ref{fig4}. The required total energy for electrons with energies above 1 GeV is $2\times10^{48}$ erg, for a source distance of 2 kpc. Although this distance is uncertain, this energy is only $\sim 0.2$\% of the typical kinetic energy available in SNRs. Therefore, the total energy required would still be reasonable for a wide range of reasonable distances. The particle energy cutoff found is $15$ GeV, and the resulting magnetic field is $13\,\mu$G. As can be seen in Fig. \ref{fig4}, the synchrotron measurements at the highest energies are in tension with the predicted fluxes, but we note that this could be caused by the simplicity of the one-zone model used. It is also clear that the dominant contribution to the gamma rays comes from bremsstrahlung emission. For a typical particle energy of 10 GeV and a magnetic field of $13\,\mu$G, the synchrotron loss time is of the order of $10^6$ yr \citep[e.g.,][]{2010MNRAS.406.2633S}. Since this time is much larger than the age of the system, our choice of the particle distribution for the electrons is justified, as no synchrotron cooling break is expected.

\subsection{Hadronic-dominated model}
\begin{figure}
 \includegraphics[width=\linewidth]{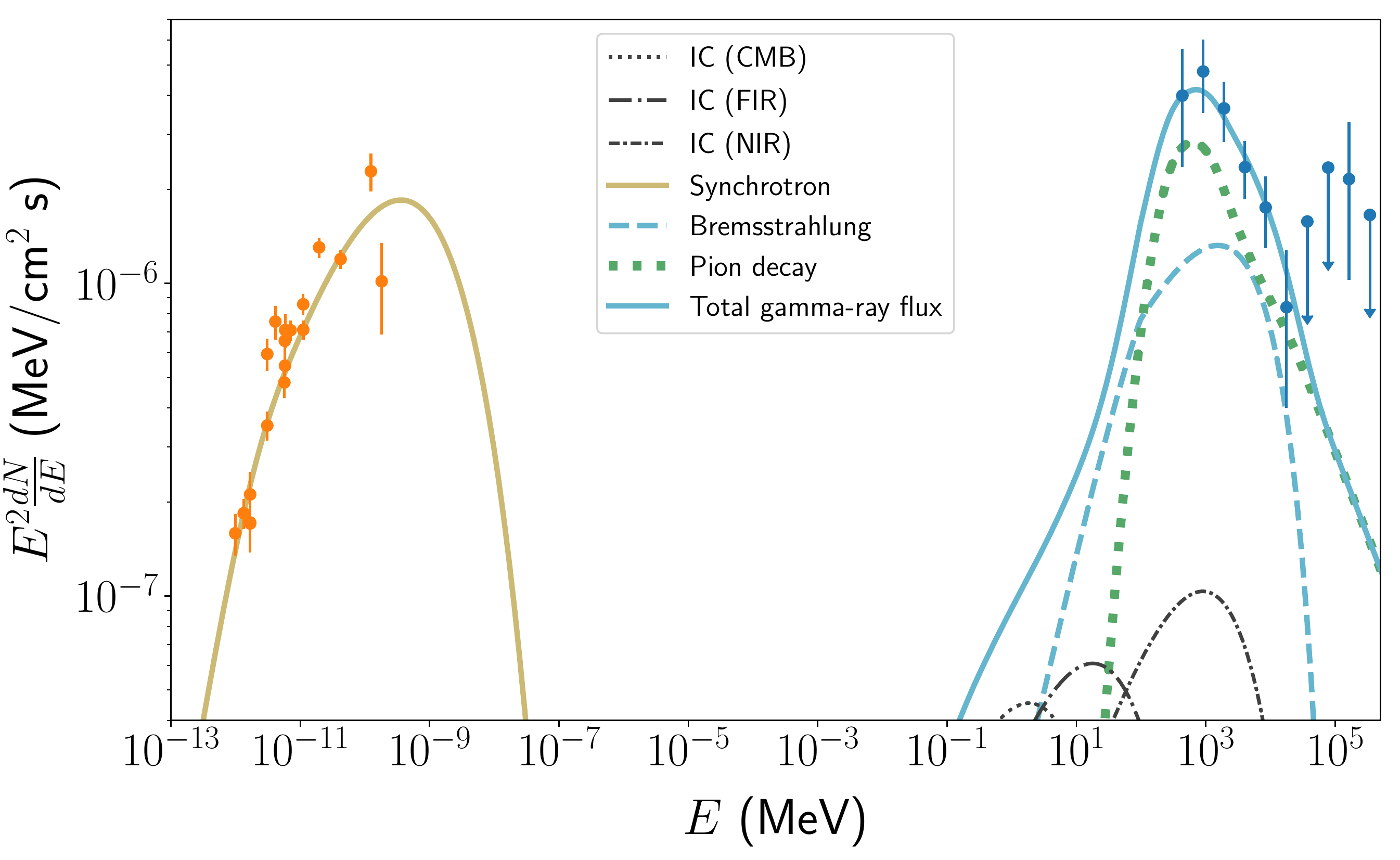}
 \caption{Hadronic-dominated model for the GeV emission from the SNR CTB 80. The gamma-ray data include both statistical and systematic errors.\label{fig5}}
\end{figure}

The fit to the multiwavelength data from CTB 80 where the GeV emission is dominated by hadronic interactions is shown in Fig. \ref{fig5}. The proton energy distribution used is a power law and the resulting particle spectral index is $2.6$. For an average particle density in the target material of $n=3$ cm$^{-3}$ \citep{1988Natur.334..229F}, a total energy content in the hadrons of $W_p=2.5\times10^{49}$ erg is required, which is $\sim 2.5$\% of the typical kinetic energy available in SNRs. The shape of the lepton distribution is the same used to obtain the leptonic model in Fig. \ref{fig4}. The total energy in the leptons and the magnetic field in this scenario were $8\times10^{47}$ erg and $21\,\mu$G, respectively. The ambient particle density used for calculation of the bremsstrahlung fluxes was also $3$ cm$^{-3}$.

Considering a volume $V$ in space occupied by the high-energy particles equal to the volume of the SNR (approximated as a sphere of radius 20 pc), the average energy density in these particles would be $W_p/V \approx 16$ eV cm$^{-3}$, which is much greater than the local energy density in Galactic cosmic rays. This implies that an enhancement of high-energy hadrons above the ISM numbers would be present at the SNR, which is consistent with the expected scenario where the SNR accelerates (or traps) cosmic rays.

As with the leptonic scenario explained above, where the gamma rays are mostly attributed to bremsstrahlung emission, a scenario where hadronic processes contribute substantially to the gamma-ray emission is also consistent with the fact that this emission is more significant at the location of the northern arm of the SNR. We recall that the IR enhancement to the north of the SNR, which is seen in Fig. \ref{fig3}, perfectly matches the radio morphology. The radio spectral variations and morphology of this region indicate that there is interaction of the shock with denser ambient medium \citep{2005A&A...440..171C}, which would then produce substantial gamma rays, and this explains the GeV morphology seen in Fig. \ref{fig1}. The shock interaction could also produce compression amplifying the magnetic field and increasing the cosmic ray energy density.

\section{Summary}
We discovered a new extended GeV source matching the location and size of the SNR CTB 80 using data from the \emph{Fermi}-LAT. The spectrum of the source is best described by a power law above an energy of 300 MeV, with a spectral index of $2.34\pm 0.09_{stat}\,\pm 0.08_{sys}$. We have shown that the GeV emission is enhanced in the northern radio arm of the SNR where the shock is believed to be interacting with material which is seen in the infrared. This fact is consistent with our spectral modeling which attributes the GeV emission to either pion decay, resulting from interactions of protons with matter, or to bremsstrahlung from high-energy electrons which also interact with matter. A more realistic scenario may include contributions from both types of particles, as shown in Fig. \ref{fig5}, since the presence of synchrotron-emitting electrons and target material inevitably results in emission of bremsstrahlung at gamma-ray energies. The required energy density in the cosmic rays is greater than the energy density in Galactic cosmic rays seen locally at Earth.

\section*{Acknowledgements}

We thank the anonymous referee for the valuable comments and for a thorough revision of the manuscript. This work was possible due to funding by Universidad de Costa Rica and its Escuela de F\'isica under grant number B8267. This research is based on observations made with NASA's Fermi Gamma-Ray Space Telescope, developed in collaboration with the U.S. Department of Energy, along with important contributions from academic institutions and partners in France, Germany, Italy, Japan, Sweden and the U.S.

\section*{Data availability.} The data underlying this article are available in the Fermi Science Support Center, at https://fermi.gsfc.nasa.gov/ssc/. The derived data generated in this research will be shared on request to the corresponding author.

\bibliographystyle{mnras}
\bibliography{references}

\bsp    
\label{lastpage}

\end{document}